\documentclass{nature}
\usepackage{epsfig}
\usepackage{color}
\usepackage{amsmath}
\usepackage{lscape}
\usepackage{amssymb}

\begin{document}

\title{Localized spatially nonlinear matter waves in   atomic-molecular Bose-Einstein condensates with space-modulated nonlinearity}
\author{Yu-Qin Yao$^{1}$, Ji Li$^{2}$, Wei Han$^{3}$, Deng-Shan Wang$^{4}$, and Wu-Ming Liu$^{2\star}$}

\maketitle

\begin{affiliations}
\item
Department of
  Applied Mathematics, China Agricultural University, Beijing 100083, People's Republic of China

\item
Beijing National Laboratory for Condensed Matter Physics, Institute of Physics, Chinese Academy of Sciences, Beijing 100190, People's Republic of  China

\item
Key Laboratory of Time and Frequency Primary Standards,
National Time Service Center, Chinese Academy of Sciences, Xi¡¯an 710600, People's Republic of China

\item
School of Science, Beijing Information Science and Technology University, Beijing 100192,  People's Republic of China


$^\star$e-mail: wliu@iphy.ac.cn

\end{affiliations}

\begin{abstract}

The intrinsic nonlinearity is the most remarkable characteristic of the Bose-Einstein condensates (BECs) systems.  Many studies have been done on atomic BECs with  time- and space- modulated nonlinearities, while there is few work considering the  atomic-molecular BECs with space-modulated nonlinearities. Here,
we obtain  two kinds of  Jacobi  elliptic  solutions  and  a family of rational  solutions of the  atomic-molecular BECs with trapping potential and space-modulated nonlinearity and consider the effect of three-body interaction on the localized matter wave solutions.
The topological properties of the localized nonlinear matter wave  for no coupling are analysed:
 the parity of nonlinear matter wave functions depends only on the principal quantum number $n$, and  the numbers of the density packets for each quantum state depend on both the principal quantum number $n$  and the secondary quantum number $l$. When the coupling is not zero,
 the localized nonlinear matter waves given by the rational function, their topological properties  are independent of the principal quantum number $n$,   only depend  on  the secondary quantum number $l$.
    The   Raman detuning   and the chemical potential  can change the number  and the shape of the
density packets.  The stability of the Jacobi  elliptic solutions depends on the principal quantum number  $n$,  while the stability of the rational solutions  depends on the chemical potential  and Raman detuning.

\end{abstract}

It is known that  the precise control of untracold atomic systems have brought the realization of Bose-Einstein condensates (BECs) and  Fermi gases.
An important challenge is to produce and control more complicated molecular systems because of their potential applications for the tests of fundamental physics and for the drifts of fundamental constants.
 To date, several atomic-molecular conversion schemes have been provided \cite{car,kes,kx,eh,add1,ead,rw,kmj,ph1,ph2}. Among them, Feshbach resonance \cite{add1,ead} and photoassociation \cite{rw,kmj} are two main techniques to produce cole molecules from an atomic BECs.  In real experiment, the cold molecules can be produced from a Fermi gas of atoms \cite{mgr,jc} or an atomic BECs based on Feshbach resonance, Raman  photoassociation or stimulated Raman adiabatic passage \cite{et,rad,tv}.  For example,   a two-photon stimulated Raman transition in a $^{87}Rb $ BECs has been used to produce $^{87}Rb_{2}$ molecules in a single rotational-vibrational state \cite{wrf}, where the input Raman laser pulse couples the molecular levels and reduces spontaneous emission.
There is a nonlinear resonant transfer between atoms and molecules, as well as term proportional to the densities in the coupled   atomic-molecular BECs.
This type of soliton solutions have been studied in the nonlinear optics \cite{mak,hh} and in the problem of the self-localization of impurity atoms BECs \cite{ks}.
The parametric solitons have been investigated in Ref.\cite{tgv}.
The coherent dynamics of this coupled atomic-molecular BECs have also been studied, which shows
very rich behaviors, such as exact dark states solution \cite{kw,rd}, crystallized and amorphous vortices \cite{cfl}, Rabi oscillations \cite{ai} and so on.

The intrinsic nonlinearity is the most remarkable characteristic of the BECs systems. In the past years,
many interesting experiments, for example,  the sonic-analogue of black holes, could be explored with spatial modulation of the interatomic interaction on short length scales. In Ref. \cite{Fedichev,Theis}, a promising technique (optical Feshbach resonance) is  proposed to control the scattering length. With the development of this topic, a successful control of a magnetic Feshbach resonance of alkali-metal atoms was illustrated in Ref.\cite{bauer}.
 In Ref.\cite{ry}, submicron control of the scattering  length has been demonstrated by applying a pulsed optical standing wave
 to a BECs of ytterbium $(^{174}Yb)$ atoms.
In recent research, the  Nonlinear Schr$\ddot{o}$dinger equation (NLSE) or the Gross-Pitaevskii equation (GPE) with spatially dependent cubic and quintic nonlinearities can be applied to the pulse propagation on optical fiber \cite{dm}, photonic crystals \cite{ys}, and the study of BECs \cite{ag,cw1}. The wide localized soliton solutions, the wide
 vector solutions, the dark soliton solutions and so on have been worked out \cite{lea,JBB,ke1,ke2}.  The localized nonlinear waves in quasi-two-dimensional BECs with spatially modulated nonlinearity and in two-component BECs with time- and space- modulated nonlinearities are constructed \cite{ds1,ds2,bam}.
 However, there is few work considering the two-dimensional atomic-molecular BECs with space-modulated nonlinearities.


 In this paper, we investigate the nonlinear matter waves in the  two-dimensional atomic-molecular Bose-Einstein condensates with space-modulated nonlinearities, which can be described by
 the  coupled GP equations with space-modulated nonlinearities. We work out three kinds of localized nonlinear wave solutions for both the attractive  spatially inhomogeneous interactions and the repulsive ones by using the  similarity transformation \cite{ata}.
Our results show that the topological properties of the localized nonlinear matter waves given by the Jacobi  elliptic  function can be described by the principal quantum number $n$ and the secondary quantum number $l$, while the topological properties of the localized nonlinear matter waves given by the rational function   are independent of the principal quantum number $n$,   only depend  on  the secondary quantum number $l$.
  The Jacobi  elliptic  solution   is linearly stable only for the principal quantum number $n=1$, while the stability of the rational form solutions  depends on the chemical potential and Raman detuning.

\section*{Results}
\subsection{The coupled Gross-Pitaevskii equation with space-modulated nonlinearity.}
In real experiment, the coherent free-bound stimulated Raman transition can cause atomic BECs of $^{87}Rb $ to produce a molecular BECs of $^{87}Rb_{2}$.
 If the molecular spontaneous emission and the light shift effect can be ignored \cite{mc,mg},
according to the mean field theory,
 the coupled  atomic-molecular BECs \cite{mg,sjw,it} with three-body interaction term can be written as
\begin{equation}
\label{eqns:eq0}
       \begin{array}{ll}
       i\hbar\frac{\partial \Psi_a}{\partial t}=(-\frac{\hbar^2\nabla^2}{2M_a}+g_a|\Psi_a|^2
       +g_{am}|\Psi_m|^2+\gamma_a|\Psi_a|^4+V^{(a)}_{ext})\Psi_a+\sqrt{2}\chi \Psi_a^*\Psi_m, \\
       i\hbar\frac{\partial \Psi_m}{\partial t}=(-\frac{\hbar^2\nabla^2 }{2M_m}+g_{am}|\Psi_a|^2
       +g_{m}|\Psi_m|^2+\gamma_m|\Psi_m|^4+V^{(m)}_{ext})\Psi_m+\frac{\chi}{\sqrt{2}} \Psi_a^2+\varepsilon\Psi_m,
       \end{array}
   \end{equation}
where $\nabla^2=\frac{\partial^2}{\partial
x^2}+\frac{\partial^2}{\partial y^2},$  $\Psi_i (i=a,m)$  denotes the
macroscopic wave function of atomic condensate and molecular
condensate respectively, $g_a=\frac{4\pi \hbar^2a_a}{M_a},
g_m=\frac{4\pi \hbar^2a_m}{M_m}$ , $g_{am}=\frac{4\pi
\hbar^2a_{am}}{M_aM_m/(M_a+M_m)}$ , $\gamma_a$ and  $\gamma_m$ represent respectively the cubic and quintic nonlinearity the strengths of
interaction, $V^{(i)}_{ext}=\frac{M_i \omega^2 (x^2+y^2)}{2} ~(i=a,m)$ are the trapping potentials, $M_a(M_m)$ is the mass of atomic (molecule),  $\chi$ is the parametric coupling coefficient which
describes the conversions of atoms into molecules due to stimulated
Raman transitions. The parameter $\varepsilon$ characterizes Raman
detuning for a two photon resonance \cite{wrf,mc,djh}.
 Integrating along the transverse coordinates, the above equations for the wave
functions $\Psi_i (i=a,m)$  in dimensionless form can be written as the coupled GP equations
\begin{equation}
\label{eqns:eq}
       \begin{array}{ll}
       i\frac{\partial \Psi_a}{\partial t}=(-\frac{\nabla^2 }{2}+g_a|\Psi_a|^2
       +g_{am}|\Psi_m|^2+\gamma_a|\Psi_a|^4+\frac{ \omega^2 (x^2+y^2)}{2})\Psi_a+\sqrt{2}\chi \Psi_a^*\Psi_m, \\
       i\frac{\partial \Psi_m}{\partial t}=(-\frac{\nabla^2 }{4}+g_{am}|\Psi_a|^2
       +g_{m}|\Psi_m|^2+\gamma_m|\Psi_m|^4+\omega^2 (x^2+y^2))\Psi_m+\frac{\chi}{\sqrt{2}} \Psi_a^2+\varepsilon\Psi_m,
       \end{array}
   \end{equation}
 The unit of length, time and energy correspond to $\sqrt{\hbar/(m \omega)}(\approx 1.07 u m),~\omega^{-1}(\approx 1.6 \times 10^{-3}s)$ and $\hbar\omega$, respectively. In this paper, we use the parameters of atomic-molecular BECs of $ ^{87}$Rb system with $M_m=2M_a=2m ~ (m=144.42\times10^{-27}Kg),$ $g_m=2g_a ~(a_a=101.8a_B),$  where $a_B$ is the Bohr radius.

Now we consider the spatially localized stationary solution
$\Psi_a=\phi_a(x,y)e^{-i\mu_at},~\Psi_m=\phi_m(x,y)e^{-i\mu_mt}$ of
 (\eqref{eqns:eq}) with $\phi_i(x,y) ~(i=a,m)$ being a real function for $lim_{|x|,|y|\rightarrow\infty}\phi_i(x,y)=0$. This maps (\ref{eqns:eq}) into the  following coupled equations
 \begin{equation}
\label{eqns:eq1}
       \begin{array}{ll}
      \frac{\partial^2 \phi_a}{\partial x^2}+ \frac{\partial^2 \phi_a}{\partial y^2}-2g_a\phi_a^3-
      2g_{am}\phi_m^2\phi_a-2\gamma_a\phi_a^5-\omega^2(x^2+y^2)\phi_a+2\mu_a\phi_a+2\sqrt2\chi \phi_a\phi_m=0 , \\
      -\frac{\partial^2 \phi_m}{\partial x^2}- \frac{\partial^2 \phi_m}{\partial y^2}+4g_m\phi_m^3+4g_{am}\phi_m\phi_a^2+4\gamma_m\phi_m^5+\omega^2(x^2+y^2)\phi_m-8\mu_a\phi_m
      +4\epsilon \phi_m-2\sqrt2\chi\phi_a^2=0,
       \end{array}
   \end{equation}
where $\mu_a,~\mu_m$ are chemical potentials.
 In order to solve the above
equations, we take the similarity transformation
\begin{equation}
\label{eqns:st} \phi_a=\beta_a(x,y)U(X(x,y)),~\phi_m=\beta_m
V(X(x,y)),
\end{equation}
 to transform  (\ref{eqns:eq1}) to the ordinary differential
equations (ODEs)
 \begin{equation}
\label{eqns:ode}
       \begin{array}{ll}
   U_{XX}+b_{11}U^3+b_{12}UV^2+b_{13}UV+b_{14}U^5=0 , \\
    V_{XX}+b_{21}U^2V+b_{22}V^3+b_{23}U^2+b_{24}V^5=0,
       \end{array}
   \end{equation}
where $b_{ij},i=1,2,~j=1,2,3$ are constants. Substituting
(\ref{eqns:st}) into (\ref{eqns:eq1}) and letting $U(X),~V(X)$ to
satisfy (\ref{eqns:ode}), we obtain a set of partial differential
equations (PDEs). Solving this set of PDEs, we have
\begin{equation}
\label{eqns:g}\begin{array}{ll}
g_a=\frac{ - \omega b_{11} }{2F^2 } e^{-\omega (x^2+4xy+y^2)} ,~
g_m=\frac{ - \omega b_{22} }{4F^2 } e^{-\omega (x^2+4xy+y^2)} ,~
g_{am}=\frac{ - \omega b_{12} }{2F^2 }e^{-\omega (x^2+4xy+y^2)},\\ \gamma_a=\frac{ - \omega b_{14} }{2F^4 } e^{-\omega (x^2+6xy+y^2)} ,~
\gamma_m=\frac{ - \omega b_{24} }{2F^4 } e^{-\omega (x^2+6xy+y^2)} ,~\\F=c_1KummerM(\frac{\omega-\mu_a}{2\omega},\frac{3}{2},\frac{\omega(y-x)^2}{2})(y-x)+
c_2KummerU(\frac{\omega-\mu_a}{2\omega},\frac{3}{2},\frac{\omega(y-x)^2}{2})(y-x),\\
 X=\frac{1}{2}\sqrt{\pi}erf(\frac{\sqrt{2\omega}(x+y)}{2}),~\varepsilon=\mu_m-\frac{1}{2}\mu_a,
   \end{array} \end{equation}
where $erf(s)=\frac{2}{\sqrt{\pi}}\int_0^se^{-\tau^2}d\tau$ is
called an error function, $KummerM(\frac{\omega-\mu_a}{2\omega},\frac{3}{2},\frac{\omega(y-x)^2}{2})(y-x)$ and $KummerU(\frac{\omega-\mu_a}{2\omega},\frac{3}{2},\frac{\omega(y-x)^2}{2})(y-x)$ are solutions of the
 the ordinary differential
equation
\begin{equation}
\label{eqns:F}
F_{YY}(Y)-\omega YF_Y(Y)+\mu_aF(Y)=0,
 \end{equation}
where $Y=y-x.$  Specially, when $\frac{\omega-\mu_a}{\omega}=3,$
the KummerM function can be simplified as exponential function $e^{\frac{\omega(y-x)^{2}}{2}}$. In this case, the interactions become as
$$g_a=\frac{ - \omega b_{11} }{2c_{1}^2 (y-x)^{2}} e^{-\omega (x^2+xy+y^2)} ,~
g_m=\frac{ - \omega b_{22} }{4c_{1}^2 (y-x)^{2}} e^{-\omega (x^2+xy+y^2)} ,~
g_{am}=\frac{ - \omega b_{12} }{2c_{1}^2(y-x)^{2} } e^{-\omega (x^2+xy+y^2)},$$
which is experimentally feasible due to the flexible and precise control of the scattering lengths achievable in BECs with magnetically tuning
the Feshbach resonances\cite{add1,ead,ry}.

\subsection{Rational solution of the atomic-molecular BECs  with three-body interaction.}
When the coupling $\chi=0,$ (\ref{eqns:st}) and (\ref{eqns:ode}) gives  the rational formal
solution of (\ref{eqns:eq})
\begin{equation}
\label{eqns:srt}\begin{array}{ll}
\Psi_a=\frac{6\sqrt{(g_{11}g_{22}-2g_{12}^{2})(g_{12}-g_{22})}F}{\sqrt{3(g_{11}g_{22}-2g_{12}^{2})^2(c-\frac{\sqrt{\pi}}{2}erf(\frac{\sqrt{2\omega}}{2}(x+y))^2+
4g_{14}(g_{12}-g_{22})^2}}e^{\omega xy}e^{-i\mu_a t},\\
\Psi_m=\frac{6\sqrt{(2g_{12}-g_{11})(g_{11}g_{22}-2g_{12}^{2})(g_{12}-g_{22})}F}{\sqrt{3(g_{12}-g_{22})((g_{11}g_{22}-2g_{12}^{2})^2(c-\frac{\sqrt{\pi}}{2}erf(\frac{\sqrt{2\omega}}{2}(x+y))^2+
4g_{14}(g_{12}-g_{22})^2})}e^{\omega xy}e^{-i\mu_a t},\\
   \end{array} \end{equation}
   where $c$ is arbitrary constant and $F$ is given in (\ref{eqns:g}).

 In order to investigate the topological properties of the exact spatially localized stationary solution  (\ref{eqns:srt}), we plot their density distributions.
 In Fig.1, it can be observed  that the energy packets are striped distribution, and the number of the energy stripes increases with the chemical potential $\mu_a$ when   $\varepsilon$ is fixed.
  It can also  be seen that
  some zero points appear on the middle density stripe along line $y=x$ when the number of the density stripes is odd.

\subsection{Jacobi elliptic function solution.}

When the three-body  effect  is very weak and
 the coupling $\chi=0,$ we have the following exact solutions of
(\ref{eqns:ode}),
\begin{equation}
\label{eqns:s2}\begin{array}{ll}
U(X)=\sqrt{\theta_1}c_0cn(c_0X,\frac{\sqrt{2}}{2}),\\
V(X)=\sqrt{\theta_2}c_0cn(c_0X,\frac{\sqrt{2}}{2})
   \end{array} \end{equation}
   or
\begin{equation}
\label{eqns:s3}\begin{array}{ll}
U(X)=\sqrt{\frac{\theta_1}{2}}d_0sd(d_0X,\frac{\sqrt{2}}{2}),\\
V(X)=\sqrt{\frac{\theta_2}{2}}d_0sd(d_0X,\frac{\sqrt{2}}{2})
   \end{array} \end{equation}
where $c_0,~d_0$ are arbitrary constants, $\theta_1=\frac{b_{22}-b_{12}}{b_{11}b_{22}-b_{12}b_{21}},~\theta_2=\frac{b_{11}-b_{21}}{b_{11}b_{22}-b_{12}b_{21}}$ and $cn,~sd=sn/dn$ are Jacobi elliptic functions.
When imposing the bounded condition $lim_{|x|,|y|\rightarrow\infty}\phi_i(x,y)=0$, we have $c_0=\frac{2(2N-1)K(\frac{\sqrt{2}}{2})}{\sqrt{\pi}}$ and $d_0=\frac{4N K(\frac{\sqrt{2}}{2})}{\sqrt{\pi}}$,
where $N$ is a natural number and $K(\frac{\sqrt{2}}{2})=\int_0^{\pi/2}[1-\frac{1}{2} sin^2\xi]^{-\frac{1}{2}}d\xi.$

From Eqs.(\ref{eqns:st}), (\ref{eqns:g}),  (\ref{eqns:s2}) and (\ref{eqns:s3}), we obtain the Jacobi elliptic function solutions  for the atomic-molecular BEC (\ref{eqns:eq})
\begin{equation}
\label{eqns:cns}\begin{array}{ll}
\Psi_a=\frac{2n\sqrt{\theta_1}K(\frac{\sqrt{2}}{2})Fe^{\omega xy}}{\sqrt{\pi}}cn(nK(\frac{\sqrt{2}}{2})erf(\frac{\sqrt{2\omega}}{2}(x+y)), \frac{\sqrt{2}}{2})e^{-i\mu_a t},\\
\Psi_m=\frac{2n\sqrt{\theta_2}K(\frac{\sqrt{2}}{2})Fe^{\omega xy}}
{\sqrt{\pi}}cn(nK(\frac{\sqrt{2}}{2})erf(\frac{\sqrt{2\omega}}{2}(x+y)), \frac{\sqrt{2}}{2})e^{-i\mu_m t},~(n=2N-1),
 \end{array} \end{equation}
 or
 \begin{equation}
\label{eqns:ds}\begin{array}{ll}
\Psi_a=\frac{\sqrt{\theta_1}2n K(\frac{\sqrt{2}}{2})Fe^{\omega xy}}{\sqrt{\pi}}sd(nK(\frac{\sqrt{2}}{2})erf(\frac{\sqrt{2\omega}}{2}(x+y)), \frac{\sqrt{2}}{2})e^{-i\mu_a t},\\
\Psi_m=\frac{\sqrt{\theta_2}2nK(\frac{\sqrt{2}}{2})Fe^{\omega  xy}}
{\sqrt{\pi}}sd(nK(\frac{\sqrt{2}}{2})erf(\frac{\sqrt{2\omega}}{2}(x+y)), \frac{\sqrt{2}}{2})e^{-i\mu_m t},~(n=2N).
 \end{array} \end{equation}

Here we discuss the existence regions of the spatially localized stationary solution  (\ref{eqns:cns}) and (\ref{eqns:ds}) by assuming the two constraint conditions
$\theta_1> 0$ and $\theta_2> 0$. We have the eight cases of parameters $b_{11},~b_{12}$ and $b_{22}$. According to the real  experiment, we consider the following two cases:

(1) $b_{22}<b_{11}<0$ and $\sqrt{2}b_{12}>\sqrt{b_{11}b_{22}}$.

  (2)  $b_{22}> b_{11}> 0$ and $-\sqrt{b_{11}b_{22}}<\sqrt{2}b_{12}<b_{11}$.

These correspond to two cases of the intercomponent interaction parameters $g_{a},~g_{m}$ and $g_{am}$:

(a) $g_{m}>g_{a}>0$ and $\sqrt{2}g_{am}<\sqrt{g_{a}g_{m}}$.

  (b) $g_{m}<g_{a}<0$ and $\sqrt{g_{a}g_{m}}>\sqrt{2}g_{am}>g_a$.

These are the regions that the exact spatially localized stationary solutions  (\ref{eqns:cns}) and (\ref{eqns:ds}) exist.
 Now we only consider case (b), which denotes two
self-attractive atom-atom interactions, two self-attractive molecular-molecular interactions, and attractive and repulsive atomic-molecular interactions.
The other cases can be analysed in the same way.

In the following,  we will see that the integer $n$ and the number of the zero points of function $F$ which equals to that of the KummerU and KummerM functions determine the topological properties of the atom and molecular packets, so we call $n$ and $l$ as the  principal quantum number and the  secondary quantum number, respectively.
In order to investigate the topological properties of the exact spatially localized stationary solution  (\ref{eqns:cns}) and (\ref{eqns:ds}), we plot their density distributions  by manipulating the principal quantum number $n$  when the  secondary quantum number $l$ is fixed.
In Fig.2, we analyse the atomic BEC when  the secondary quantum number
$l$ is fixed and the principal quantum number $n$  is modulated. It is easy to see that the number of density packets  for each quantum states is equal to $2n$. And the number of density packets on each quantum  states  increases two by two when  the  principal quantum number $n$ increases.  The properties of the molecular BEC are similar to that of the atomic BEC.
In Fig.3, we analyse the interactions of the atomic BEC   and the molecular BEC  when the secondary quantum number $l$ is fixed. It is shown that the interaction is stronger when $N=1$ and   becomes   weaker with the increasing of $N$.

When the principal quantum number $n$ is fixed, we can adjust the secondary quantum number $l$  to observe the properties of the  atomic-molecular BEC.
Fig.4  demonstrates the density distributions of atomic-molecular BEC for different  secondary quantum number $l$. It is easy to find that the number of energy packets increases when $l$ increases, and the number of the nodes for each quantum state equals to the secondary quantum number $l$. And  some zero points appear on the middle density packets  along the line $y=x$ when the number of the  secondary quantum number $l$ is even.
 Fig.5 demonstrates the interaction of the atomic BEC and molecular BEC when the principal quantum number $n$ is fixed.
It is shown that  the number of the  atomic-molecular pair is the function of the
  secondary quantum number $l$ and  some zero points appear on the middle atomic-molecular pair  along the line $y=x$ when the number of the  secondary quantum number $l$ is even.

 Now we analyse the effect of  Raman detuning $\varepsilon$ for  the  atomic-molecular BEC.  From Fig.6, we can see that when $\varepsilon <  \mu_a$ and $\varepsilon$ is fixed,  the number of the  density packets increases one by one with the increasing of the chemical potential $\mu_a$.
 When $\varepsilon\geq\mu_a$, there is  only one density packets for each quantum states.
 The  absolute of  $\varepsilon-\mu_a$  affect the shape of the energy packets: when the absolute of  $\varepsilon-\mu_a$ is small, the shape of the density packet is like circle, and  when the absolute of  $\varepsilon-\mu_a$ is larger, the shape of the density packet becomes narrow and long.

\subsection{Rational formal solution.}

When the three-body effect is very weak and the coupling $\chi\neq 0$,
(\ref{eqns:st}) and (\ref{eqns:ode}) also gives the rational formal
solution of (\ref{eqns:eq})

\begin{equation}
\label{eqns:sr}\begin{array}{ll}
\Psi_a=\frac{-12 \sqrt{2}b_{13}F}{b_{13}^2(2c-\sqrt{\pi}erf(\frac{\sqrt{2\omega}}{2}(x+y))^2+27b_{11}}e^{\omega xy}e^{-i\mu_a t},\\
\Psi_m=\frac{-24b_{13}F}{b_{13}^2(2c-\sqrt{\pi}erf(\frac{\sqrt{2\omega}}{2}(x+y))^2+27b_{11}}e^{\omega xy}
e^{-2i\mu_a t},\\
\chi=\frac{\omega \sqrt{2}b_{13}}{8F }e^{-\omega (x^2+3xy+y^2)}.
   \end{array} \end{equation}
   where $b_{13}=b_{23},~b_{11}>0,~c$ is arbitrary constant and $F$ is given in (\ref{eqns:g}).

   In order to investigate the topological properties of the exact spatially localized stationary solution  (\ref{eqns:sr}), we plot their density distributions by adjusting the secondary quantum number $l$. The secondary quantum number $l$ is always zero  for $\varepsilon <\mu_a$, and can be taken different values for  $\varepsilon\geq\mu_a$.
 In Fig.7, it can be observed  that the energy packets are striped distribution, and the number of the energy stripes increases with the chemical potential $\mu_a$ when the secondary quantum number $l=0$ and   $\varepsilon$ is fixed.
 When the secondary quantum number $l\neq0$, there is only one energy stripe and the energy stripe becomes more narrower with the increasing of the secondary quantum number
$l$. It can also  be seen that
  some zero points appear on the middle density stripe along line $y=x$ when the number of the density stripes is odd. Fig.1 and Fig.7 show that the rational solution  (\ref{eqns:srt}) and (\ref{eqns:sr}) have similar  topological properties, which implies that three-body interaction doesn't hinder the formation of  the localized nonlinear matter wave solutions.

\subsection{Linear stability analysis.}
In the following, we analyse the linear stability of the solutions (\ref{eqns:cns}), (\ref{eqns:ds}) and (\ref{eqns:sr}) by using the linear stability
 analysis. A perturbed solution is constructed as \cite{yang,BAM1}
 \begin{equation}
\label{eqns:p}\begin{array}{ll}
\Psi_a=[\phi_a(x,y)+u_1(x,y)e^{i\lambda t}+w_1^*(x,y)e^{-i\lambda t}]e^{-i\mu_a t},\\
\Psi_m=[\phi_m(x,y)+u_2(x,y)e^{i\lambda t}+w_2^*(x,y)e^{-i\lambda t}]e^{-i\mu_m t}
   \end{array} \end{equation}
where $\mid u_1\mid\ll1,~\mid u_2\mid\ll1, ~\mid w_1\mid\ll1,~\mid w_2\mid\ll1$ are small perturbation. Substituting this perturbed solution
into  (\ref{eqns:eq}) and neglecting the higher-order terms in $ u_1,~u_2,~w_1$ and $ w_2$, we obtain the eigenvalue problem
{\footnotesize \begin{equation}\label{eqns:eigen1}
\left(\begin{array}{cccc}L_1& \sqrt{2}\chi \phi_m-g_a\phi_a^{2}&\sqrt{2}\chi \phi_a-g_{am}\phi_a\phi_m&-g_{am}\phi_a\phi_m\\g_a\phi_a^{2}-\sqrt{2}\chi \phi_m&-L_1&g_{am}\phi_a\phi_m&g_{am}\phi_a\phi_m-\sqrt{2}\chi \phi_m\\
\sqrt{2}\chi \phi_a-g_{am}\phi_a\phi_m&-g_{am}\phi_a\phi_m&L_2&-g_m\phi_m^{2}\\
g_{am}\phi_a\phi_m&g_{am}\phi_a\phi_m- \sqrt{2}\chi \phi_a&g_m\phi_m^{2}&-L_2
\end{array}\right)\left(\begin{array}
{cccc}u_1\\w_1\\u_2\\w_2\end{array}\right)=\lambda\left(\begin{array}
{cccc}u_1\\w_1\\u_2\\w_2\end{array}\right),
\end{equation}}

where $$L_1=\frac{1}{2}(\partial_x^{2}+\partial_y^{2})-2g_a\phi_a^{2}-g_{am}\phi_m^{2}-\frac{1}{2}\omega^2(x^2+y^2)+\mu_a,$$~
$$L_2=\frac{1}{4}(\partial_x^{2}+\partial_y^{2})-2g_m\phi_m^{2}-g_{am}\phi_a^2-\frac{1}{4}\omega^2(x^2+y^2)+2\mu_a-\varepsilon.$$
Numerical experiments show that the eigenvalue $\lambda$ of the eigenvalue problem  (\ref{eqns:eigen1}) is real for  $n=1$. This suggests that the localized nonlinear matter wave solution (\ref{eqns:cns}) is linearly stable for $n=1$ and solution (\ref{eqns:ds}) is unstable.
For the solution  (\ref{eqns:sr}), it can be shown that the linear stability rests on the chemical potential $\mu_a$ and the Raman detuning $\varepsilon$ (see Fig.8).

\section*{Discussion}

In this paper, we focus on the  analytic solutions of atomic-molecular BECs and the effects of the coupling $\chi$ and the Raman detuning $\varepsilon$ on the atomic-molecular BECs. The system in this report is like the one in the Ref.\cite{sjw}. Comparing to the atomic-molecular system given in the Ref.\cite{sjw}, Gupta and Dastidar have proposed a more complicated model when they study the dynamics of atomic and molecular BECs of $^{87}$Rb in a spherically symmetric trap coupled by stimulated Raman photoassociation process in the Ref.\cite{mg}. In fact, the light shift effect in Gupta and Dastidar's model almost has the same function as the Raman detuning term. So, it can be contributed to the Raman detuning term. Based on this reason, we don't consider the light shift effect and take the form of the atomic-molecular BECs system as the form in Ref.\cite{sjw}.

In  the Ref.\cite{sjw}, they show that the coherent coupling between atoms and molecules changes the situation crucially  and it is sensitive to the presence of vortices. For example, when the coupling $\chi$ is zero, each of the atoms and molecular BECs wave function forms an independent triangular vortex lattice, and a nonzero coupling $\chi$ proposes more dramatic changes. Our results show that the coupling $\chi$ can change the topological structure of the localized nonlinear wave of the atomic- molecular BECs. In the case of $\chi=0$, Fig.1-Fig.4 illustrate that the topological structures depend on  the principal quantum number $n$ and the secondary quantum number $l$, and each density packet is like a circle and oval. When $\chi \neq 0$, Fig.6 display the density packets are striped distribution and their topological structures only reply on the secondary quantum number $l$ and are independent on the principal quantum $n$.

In real experiment, spatial modulation of the interatomic interaction can be achieved. In the recent experiment\cite{ry}, the authors apply a pulsed optical standing wave to a BEC of ytterbium ($^{174}$Yb) atoms and realize the submicron control of the scattering length. The experimental phenomena is well explained by the semi-classical theory of Bohn and Julienne\cite{bohn}. In this paper, the interaction $g_a,~g_m,~g_{am}$  and the coherent coupling $\chi$
all depend on the spatial variables. Under that conditions, the stable exact solutions can be worked out for the first time.  The spatial modulation of the interaction can be realized by the above experiment, but there is no successful experiment for the spatial modulation of  the coherent coupling. We hope that our research will stimulate the further research on the spatial modulation of the atomic- molecular BECs.

It is obvious that the Raman detuning term in the atomic- molecular BECs behaves just like the chemical potential to control the system's energy.
In this paper, the results imply that $\mu_{a}-\varepsilon$ not only changes the altitude of the wave packets, but also changes the  topological structures of the nonlinear waves. When $\mu_{a}-\varepsilon\geq 0$, the number of the energy packets changes with the chemical potential  $\mu_{a}$.  When $\mu_{a}-\varepsilon < 0$, there is only one  energy packet for each quantum state.


In summary, we have worked out three kinds of  localized nonlinear matter wave solutions of the two-dimensional atomic-molecular BECs with space-modulated nonlinearity and considered the effect of three-body interaction on the localized nonlinear matter wave solutions.
Our results show that the matter wave functions given by elliptic function have even parity for the even principal quantum number and odd parity for the odd one,  the number of density
packets for each quantum state is twice of the principal quantum number $n$, and the number of  density packets increases two by two  with  the principal quantum number $n$. The number of the nodes equals to the secondary quantum number $l$.
For the nonlinear matter wave given by rational function,
 the number of the energy stripes increases with the chemical potential $\mu_a$ when the secondary quantum number $l=0$ and   $\varepsilon$ is fixed.
 When the secondary quantum number $l\neq0$, there is only one energy stripe for each quantum state and the energy stripe becomes more narrower with the increasing of the secondary quantum number $l$.
 Odd (even) secondary quantum number $l$ leads to even (odd)  number of the energy packets (stripes). Some zero points appear on the middle energy packets (stripes) along line $y=x$ for even secondary quantum number $l$.
 We also analyse the  effect of  Raman detuning $\varepsilon$ for  the atomic-molecular BECs. The value of  $\varepsilon-\mu_a$ can change the number and shape of the energy packets (stripes).
  The stability of our solutions  is analysed: the nonlinear matter wave solution (\ref{eqns:cns}) is linearly stable for the principal quantum number $n=1$, the solution (\ref{eqns:ds}) is unstable, and the stability of the solution  (\ref{eqns:sr}) rests on the chemical potential $\mu_a$ and the Raman detuning $\varepsilon$. Our results are significant to matter wave management in high-dimensional  atomic-molecular BECs.

\section*{Methods}

We use the coupled Gross-Pitaevskii equation  to describe the  atomic-molecular BECs. Taking into account the term responsible for the creation of molecules \cite{ Bradley}, the Hamiltonian is taken as
\begin{equation}
\widehat{H}=\int d^{3}r(\widehat{\Psi^{*}_a}[-\frac{h^2}{2m}\nabla^2+\frac{g_a}{2}\widehat{\Psi^{*}_a}\widehat{\Psi_a}]\widehat{\Psi_a}
+\widehat{\Psi^{*}_m}[-\frac{h^2}{4m}\nabla^2+\varepsilon+\frac{g_m}{2}\widehat{\Psi^{*}_m}\widehat{\Psi_m}]\widehat{\Psi_m}
\end{equation}
$$+g_{am}\widehat{\Psi^{*}_a}\widehat{\Psi_a}
\widehat{\Psi^{*}_m}\widehat{\Psi_m}+\frac{\chi}{\sqrt{2}}[\widehat{\Psi^{*}_m}\widehat{\Psi_a}\widehat{\Psi_a}+\widehat{\Psi_m}\widehat{\Psi^{*}_a}\widehat{\Psi^{*}_a}]).$$
First, the coupled Gross-Pitaevskii equation is decomposed into two ODEs and a number of PDEs making use of the  similarity transformation.  Then we solve these ODEs and PDEs by using some solving techniques and some special functions, such as error function, KummerU function and Jacobi elliptic function.
The final interaction parameters are altered to $g_m=2g_a,~g_{am}=\frac{1}{3} g_a$ and the chemical potential satisfies $\mu_m=2\mu_a$.



\begin{addendum}

\item [Acknowledgement]

Y. Q. Y. was supported by the NSFC under Grant No. 11301179.
W. M. L. is supported
the NSFC under grants Nos. 11434015, 61227902, 61378017,  11301179, 11271362 and 11375030, NKBRSFC under grants Nos. 2012CB821305, SKLQOQOD under grants No. KF201403, SPRPCAS under grants Nos. XDB01020300 and XDB21030300, Beijing Nova program No. Z131109000413029 and Beijing Finance Funds of Natural Science Program for Excellent
Talents No. 2014000026833ZK19.

\item [Author Contributions]

W.M.L. conceived the idea and supervised the overall research. Y.Q.Y.,  W.H. and. D.S.W
performed the computations and writing the program for pictures. J.L. performed the analyse of linear stability.  Y.Q.Y. wrote the
paper with helps from all other co-authors.

\item [Competing Interests]
The authors declare that they have no competing financial interests.

\item [Correspondence]
Correspondence and requests for materials should be addressed to Liu, Wu-Ming.
\end{addendum}

\clearpage

\newpage
\textbf{Figure 1 The density distributions $|\psi_a|^{2}$ of the atomic-molecular BEC with three-body interaction term as the function of $\varepsilon$ and  $\mu_a$ with $\omega=0.02,~ b_{11}=3,~b_{22}=12, ~b_{12}=1$.} The energy packets are striped distribution. (a)-(c) show that the number of the energy stripes  increases with chemical potential $\mu_a$ when  $\epsilon$ is fixed.  (d)-(f) illustrate that some zero points appear on the middle density stripe when the number of the density stripes is odd.

\bigskip
\textbf{Figure 2 The density distributions $|\psi_a|^{2}$ of the  atomic BEC as the function of  the  principal quantum number $n$   when   the secondary quantum number is fixed.}  The wave function $\psi_a$  take the form in Eqs.(\ref{eqns:cns}) and (\ref{eqns:ds}) with $ b_{11}=3,~b_{22}=12, ~b_{12}=1$ and $\omega=0.2$.  The number of density packets  for each quantum states is equal to $2n$. (a)-(c) show the density distributions of the odd parity wave functions (\ref{eqns:cns}) for $n=1,3,5,$ respectively. (d)-(f)  illustrate  the density distributions of the even parity wave functions (\ref{eqns:ds}) for $n=2,4,6,$ respectively. The solution displayed in figuer (a) is linear stable. The unit of the length is 1.07 $\mu m$.

\bigskip

\textbf{Figure 3 The density distributions  $|\psi_a|^{2}+|\psi_m|^{2}$ of  the atomic-molecular pair  as the function of $N$  when   the secondary quantum number $l=1$.} Here $ b_{11}=3,~b_{22}=12, ~b_{12}=1,$ and $\omega=0.2$. (a) shows the interaction of the atomic-molecular pair for $N=1$.
(b) shows the interaction of the middle atomic-molecular pair for $N=2$, and it also displays that the interactions are weaker than the interaction in (a).
 (a)-(f) illustrate that the interaction  becomes   weaker with the increasing of $N$. The unit of the length is 1.07 $\mu m$.

\textbf{Figure 4 The density distributions  $|\psi_a|^{2}$ of the  atomic-molecular BEC  as the function of the secondary quantum number $l$.}  Here $\omega=0.2,~ b_{11}=3,~b_{22}=12, ~b_{12}=1$. The number of the nodes for each quantum state equals to the secondary quantum number $l$ and some zero points appear on the middle one when the number of the density packets is odd. (a1)-(a4) show the density distributions of the odd parity wave functions (\ref{eqns:cns}) for $n=1$ and $l=1,~2,~3,~4,$ respectively. (b1)-(b4) show the density distributions of the even parity wave functions (\ref{eqns:ds}) for $n=2$ and $l=1,~2,~3,~4$, respectively. The solutions displayed in the first and third figures on the upper row are linear stable.
 The unit of the length is 1.07 $\mu m$.

\bigskip

\textbf{Figure 5 The density distributions $|\psi_a|^{2}+|\psi_m|^{2}$ of the atomic-molecular pair   as the function of the secondary quantum number $l$   with $\omega=0.2,~ b_{11}=3,~b_{22}=12, ~b_{12}=1$. } The number of the atomic-molecular pairs equals to $l+1$.   (a)-(c) show the  density profiles of the atomic-molecular pair
 for $l$ is odd. (d)-(f) show the  density profiles of the atomic-molecular pair
 for $l$ is even, and it also displays that some zero points appear on the middle one. The unit of the length is 1.07 $\mu m$.

\bigskip

\textbf{Figure 6 The effect of  Raman detuning $\varepsilon$ for  the atomic-molecular BEC.} (a), (b) and (c) show that the number of the  density packets increases with the chemical potential $\mu_a$ when  $\varepsilon <\mu_a$.  (a) and (d)  reveal that the number of the  density packets don't depends on the  chemical potential $\mu_a$  and there is only one density packet for each quantum state when  $\varepsilon \geq \mu_a$, it also show the value of $\varepsilon -\mu_a$ effects the shape of the density packet. The solution displayed in figure (b) is linear stable.
The unit of the length is 1.07 $\mu m$.

\bigskip

\textbf{Figure 7 The density distributions $|\psi_a|^{2}$ of the atomic-molecular BEC as the function of $\varepsilon$ and  $\mu_a$ with $\omega=0.02,~ b_{11}=3,~b_{22}=12, ~b_{12}=1$.} The energy packets are striped distribution. (a1)-(b3) show that the number of the energy stripes  increases with chemical potential $\mu_a$ when the secondary quantum number $l=0$. (c1)-(c3)  show that there is only one density stripe when the secondary quantum number $l\neq0$. (b1) and (c3) illustrate that some zero points appear on the middle density stripe when the number of the density stripes is odd. The solutions displayed in figures  (a1) and (b1) are linear stable.

\textbf{Figure 8 Linear stability. } Eigenvalue for different principal quantum numbers $n$ with parameters  $ b_{11}=3,~b_{22}=12, ~b_{12}=1$. (a1)-(a3) show that the exact solution (\ref{eqns:cns}) is linearly stable only for $n=1$; (b1)-(b3) show that the exact solution (\ref{eqns:ds}) is linearly unstable for all $n$; (c1)-(c3) illustrate that the solution (\ref{eqns:sr}) are linearly stable in the two group parameters $\varepsilon=0.01,~\mu_m=0.1$ and  $\varepsilon=0.1,~\mu_m=0.2$.

\begin{figure}
\begin{center}
\epsfig{file=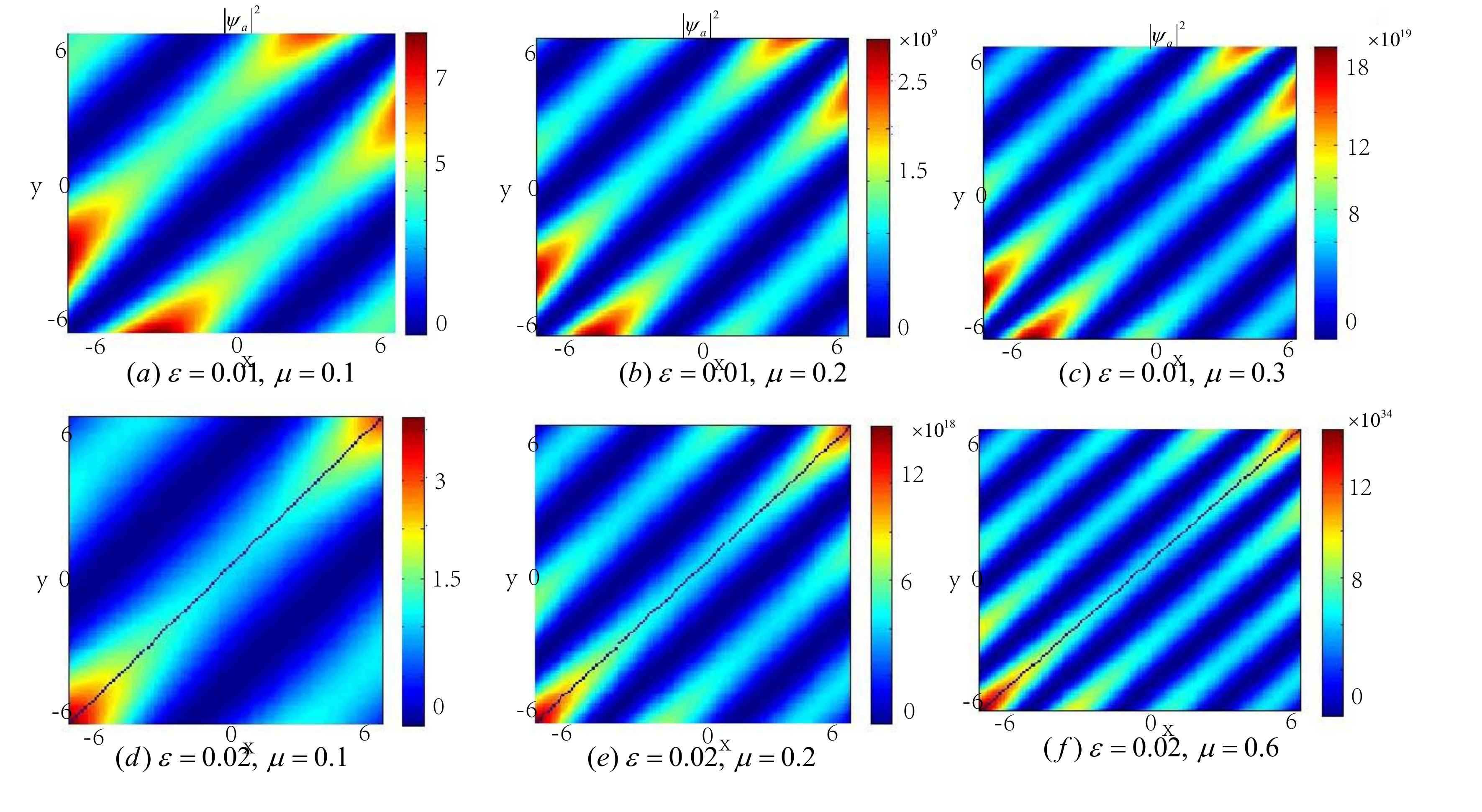,width=17cm}
\end{center}
\label{fig:TQPT}
\end{figure}

\begin{figure}
\begin{center}
\epsfig{file=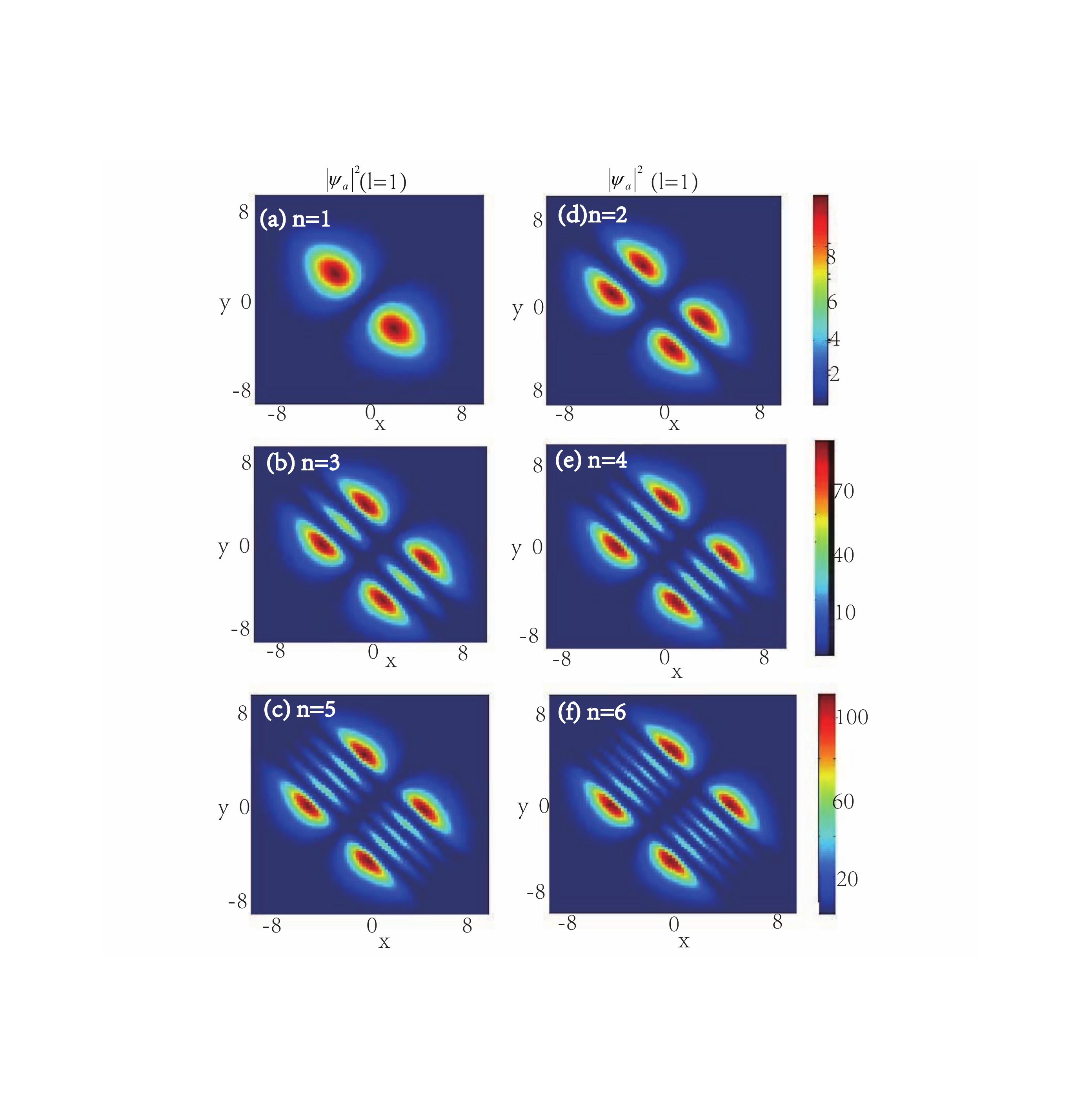,width=17cm}
\end{center}
\label{fig:TQPT}
\end{figure}

\begin{figure}
\begin{center}
\epsfig{file=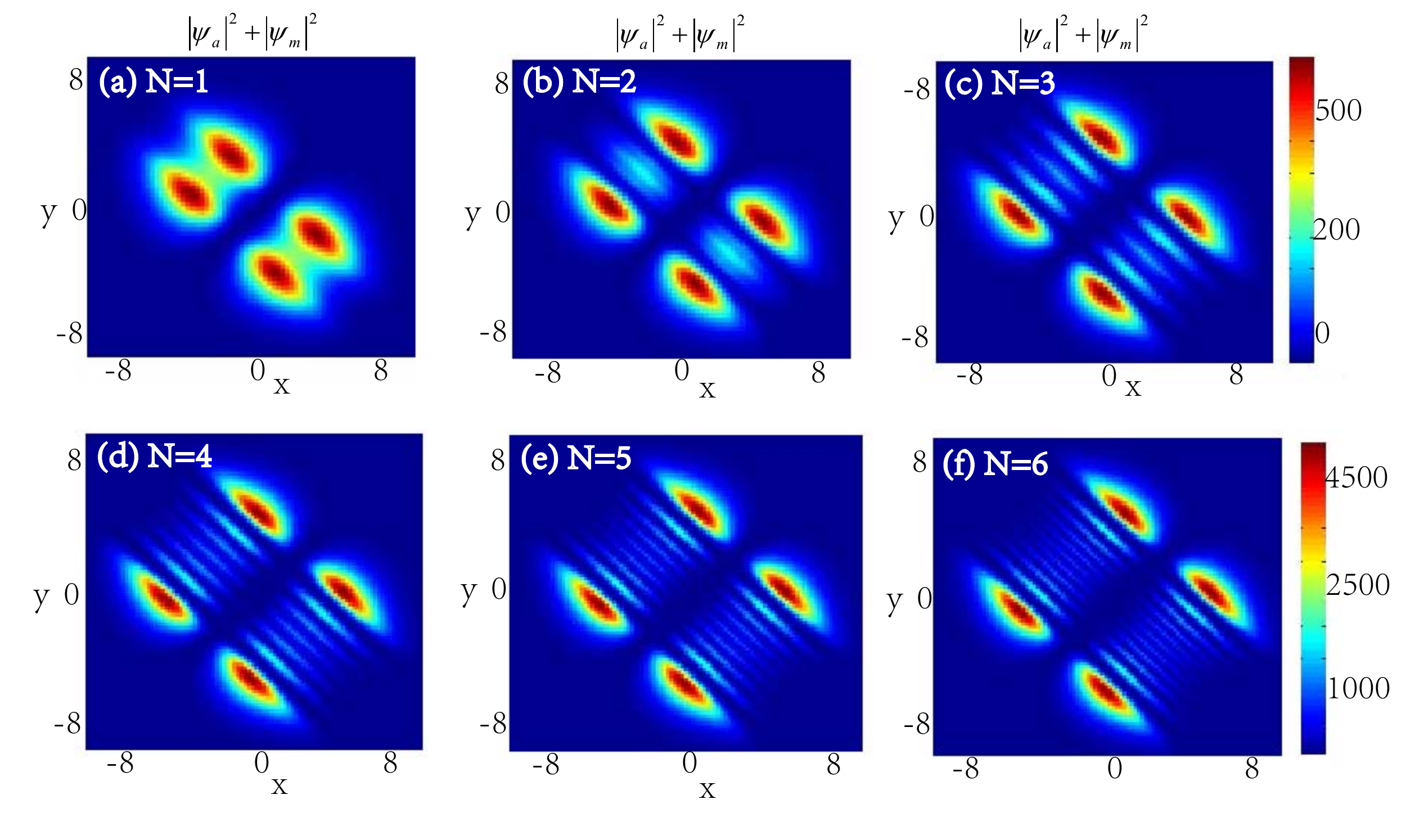,width=14cm}
\end{center}
\label{fig:FiniteT}
\end{figure}

\begin{figure}
\begin{center}
\epsfig{file=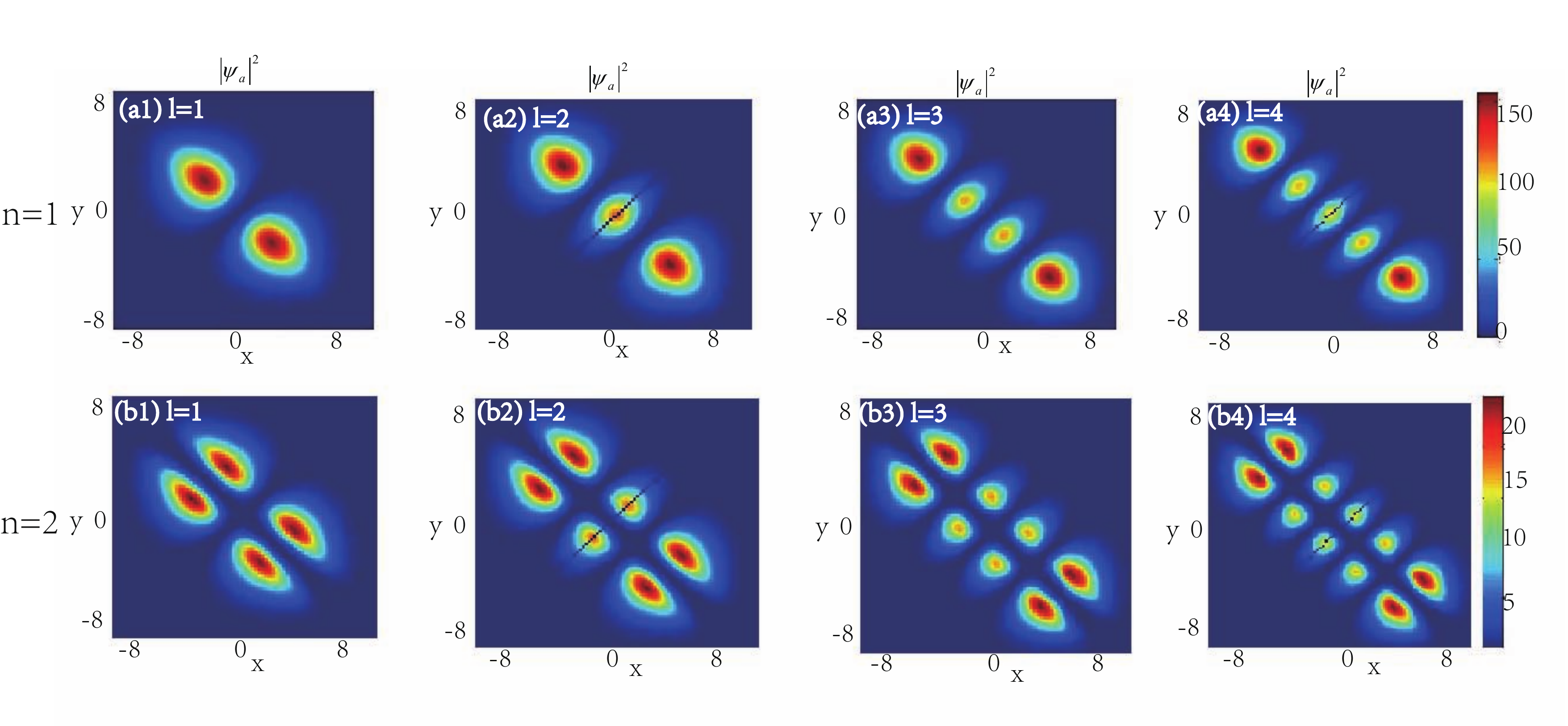,width=18cm, trim=0.5in 0.0in 0.0in 0in}
\end{center}
\label{fig:FiniteT}
\end{figure}
\clearpage

\begin{figure}
\begin{center}
\epsfig{file=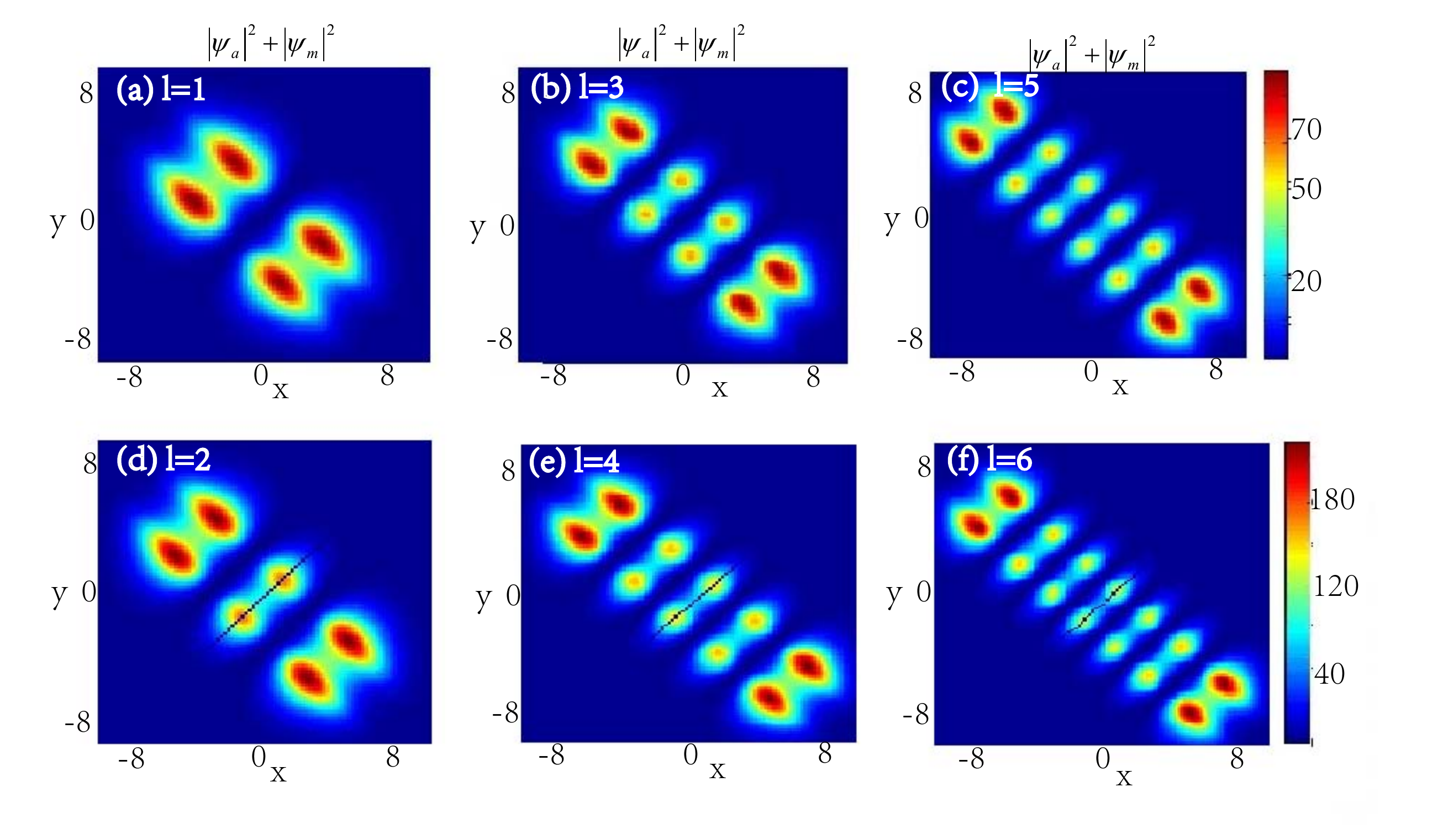,width=15cm}
\end{center}
\label{fig:FiniteT}
\end{figure}

\begin{figure}
\begin{center}
\epsfig{file=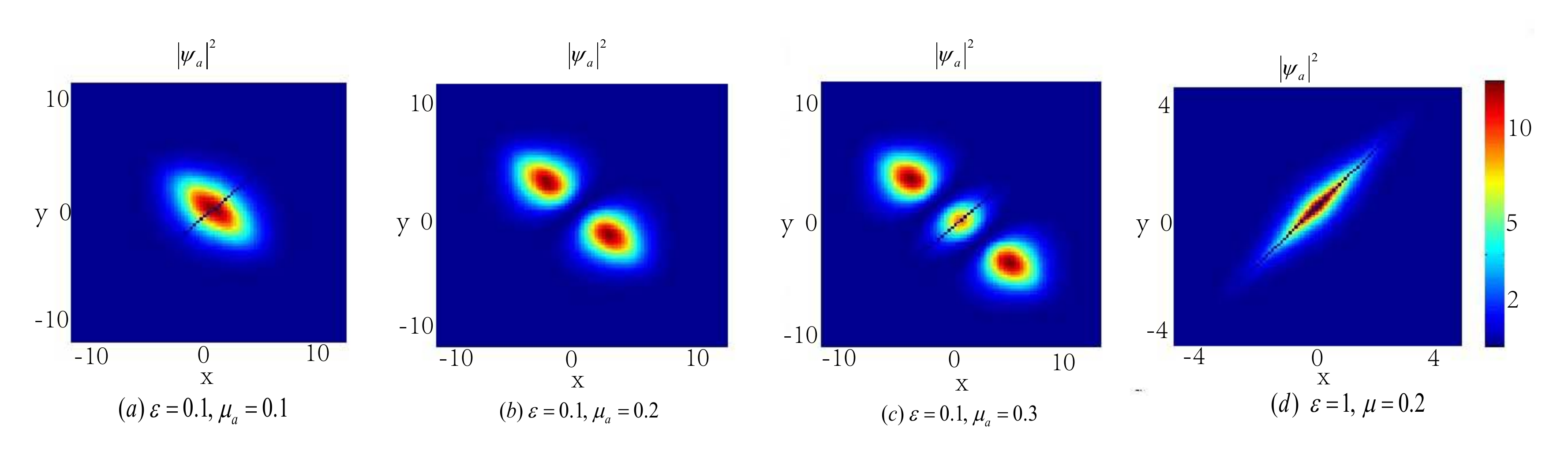,width=17cm}
\end{center}
\label{fig:ES}
\end{figure}

\begin{figure}
\begin{center}
\epsfig{file=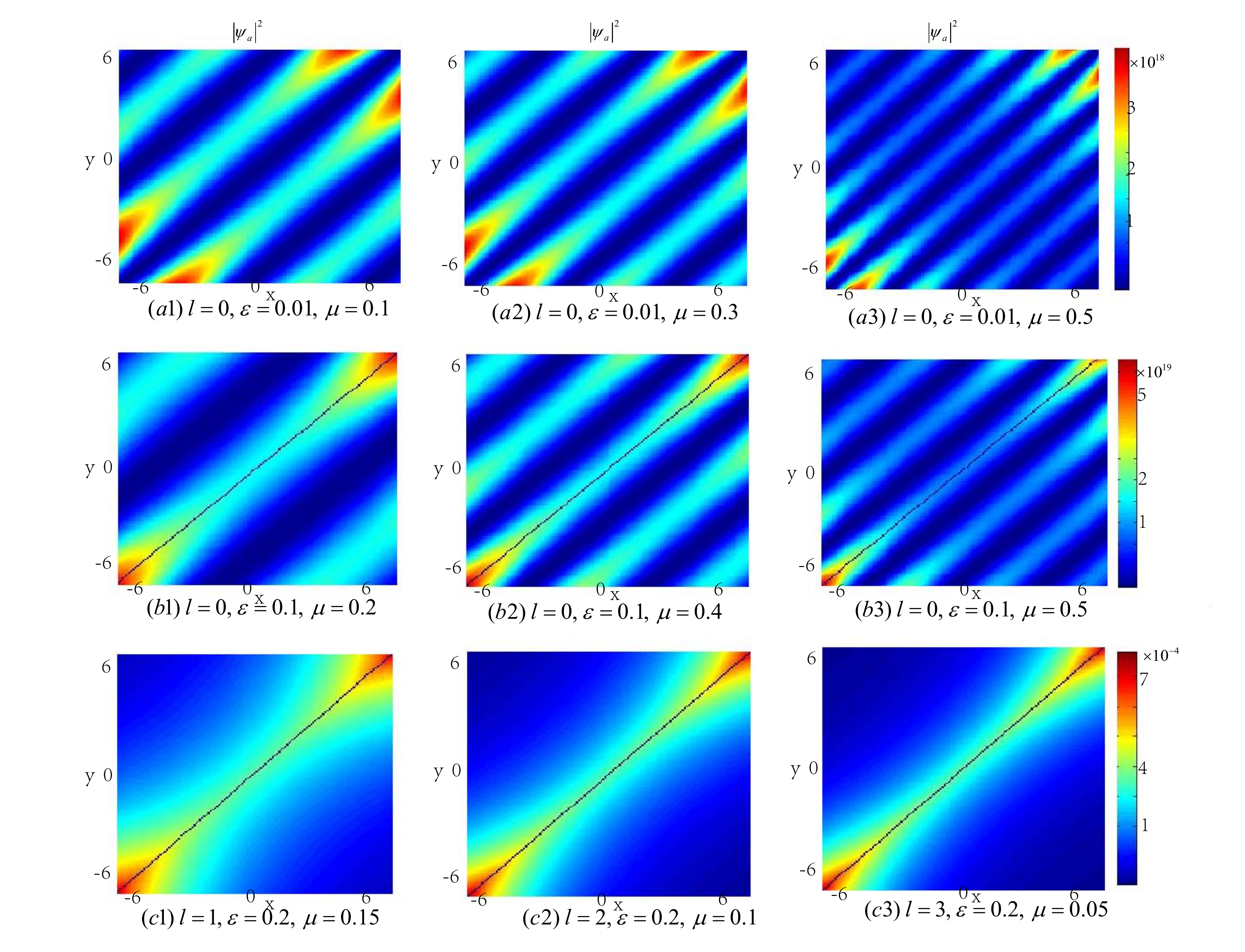,width=17cm}
\end{center}
\label{fig:ES}
\end{figure}

\begin{figure}
\begin{center}
\epsfig{file=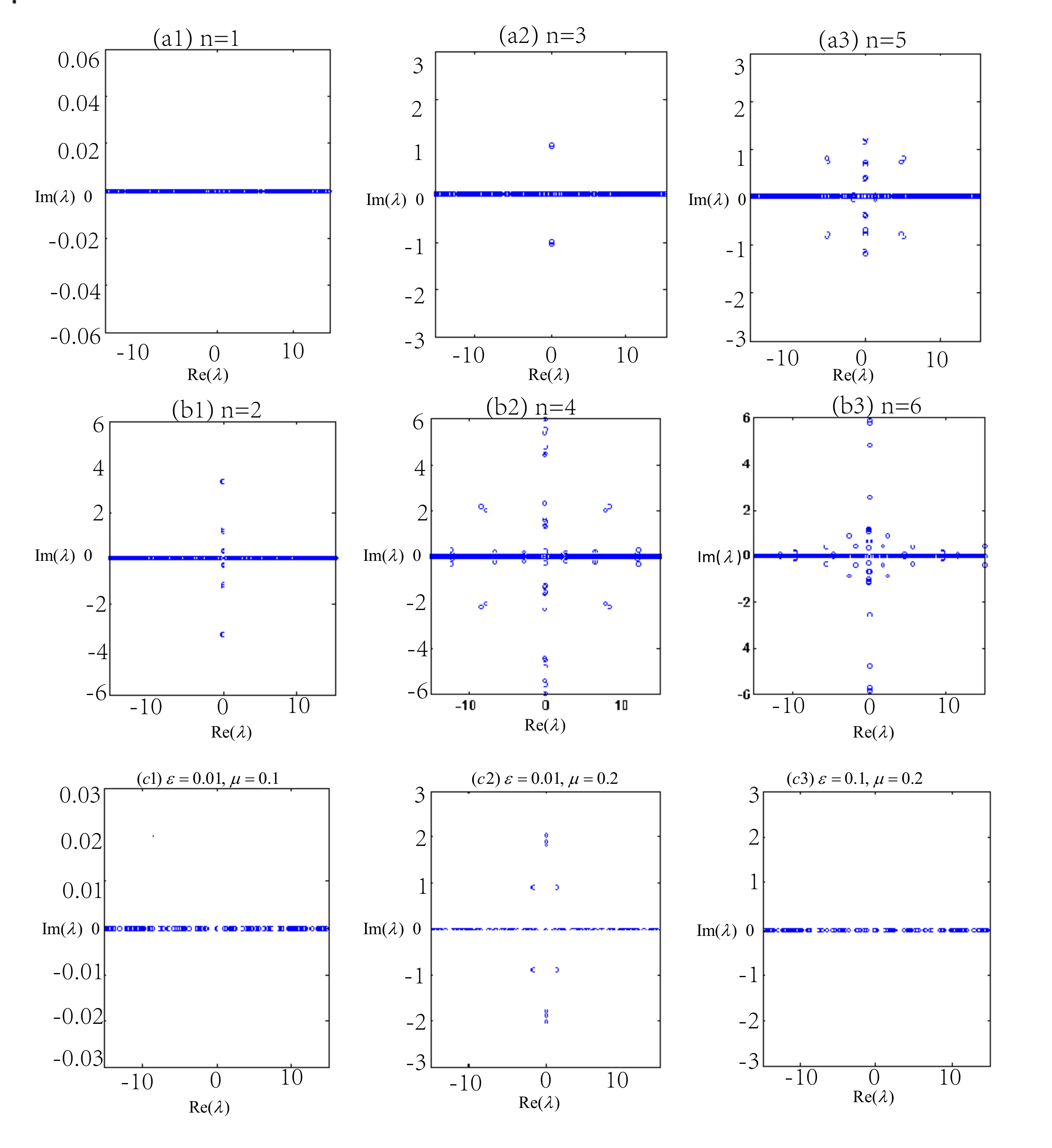,width=17cm}
\end{center}
\label{fig:ES}
\end{figure}

\end{document}